\documentclass[preprint,showpacs,preprintnumbers,amsmath,amssymb,superscriptaddress]{revtex4}

\include{ams}
\usepackage{amsmath,amssymb,amsthm}
\usepackage{dcolumn}
\usepackage{bm}
\usepackage{graphicx}

\begin{document}

\title[Field emission of
crystalline graphite]{Effect of band structure on field emission
of crystalline graphite}

\author{V L  Katkov and V A Osipov}

\address{Joint Institute for Nuclear Research, Bogoliubov Laboratory of Theoretical Physics,
141980 Dubna, Moscow region, Russia} \email{katkov@theor.jinr.ru,
osipov@theor.jinr.ru}

\begin{abstract}
The field emission of crystalline $AAA$ graphite is studied within
a simple analytical approach with account of the exact dispersion
relation near the Fermi level. The emission current is calculated
for two crystal orientations with respect to the applied electric
field. It is found that the exponent of the Fowler-Nordheim
equation remains the same while the preexponential factor is
markedly modified. For both field directions, the linear field
dependence is found in weak fields and the standard quadratic
Fowler-Nordheim behavior takes place in strong fields. A strong
dependence of the emission current from the interlayer distance is
observed. As an illustration of the method the known case of a
single-walled carbon nanotube is considered.
\end{abstract}

\pacs{79.70.+q, 81.05.Uw}

%\submitto{\JPCM}

\maketitle
\section{Introduction}

Different carbon-based structures are considered as promising
electrode material for field emission (FE) cathodes. In
particular, the field emission properties of single-walled (SWNT)
and multi-walled (MWNT) carbon nanotubes \cite{Bonard} as well as
graphite films \cite{Obraztsov} are presently under intensive
experimental and theoretical investigations. In experiment, many
factors such as inhomogeneities at the cathode surface, surface
contamination (surface adsorbates and oxides), local electric
fields and barriers, electronic structure of cathode, etc. can
drastically change FE results~\cite{modine}. In addition, these
factors vary from one experiment to another thus markedly
complicating the theoretical description. Nevertheless, the
electronic characteristics of cathodes should be equally
manifested in different experiments. For this reason, the effect
of electronic structure on the emission features of cathodes
is of definite interest. For SWNTs this problem was studied
numerically in \cite{china1, china2,china3} by using an approach
which can be called as a method of {\it independent channels}.

In this paper, we present a rather simple modification of this
method to study analytically the influence of the 3D band
structure on the field emission current (FEC) of crystalline
graphite. As an illustration, we consider the case of SWNT where
our approach allows us to reproduce the FE results obtained
in~\cite{china1}. As is well-known, the electronic structure near
the Fermi energy of the crystalline graphite markedly depends on
the weak interlayer interaction (see, e.g., \cite{Wallace,
Sloncz,McClure,Charlier}). Accordingly, the FEC in this case
should be sensitive to the specific electronic structure. In order
to show this, we consider the simplest possible modification of
graphite (hypothetical $AAA$ stacking) where the three-dimensional
energy spectrum was calculated analytically in~\cite{Charlier}.
Two possible orientations of the applied electric field (along and
normal to the graphite layers) are of our interest.

\section{FEC of opened carbon SWNT}

The emitted current density can be written as~\cite{modine,gp}
\begin{equation}\label{main}
  j^{out} = \frac{2e}{h^{3}}\int d p_x
  \int d p_y \int
  f(\varepsilon)\upsilon_g D(\varepsilon, p_x, p_y)
  d p_z.
\end{equation}
Here the field emission is directed along the $z$-axis, $e$ is the
electric charge, $h = 2\pi \hbar$ the Planck constant,
$\varepsilon$ the  energy, $\textbf{p}$ momentum,
$f(\varepsilon)=[\exp(\varepsilon/kT)+1]^{-1}$ the Fermi-Dirac
distribution function, $D(\varepsilon, p_x, p_y)$ the transmission
probability of an electron through a potential barrier, and
$\upsilon_g =
\partial\varepsilon/\partial p_z$ the group velocity.  The integrals are
over the first Brillouin zone with account of the positivity of
$\upsilon_g$. Notice that in most cases the using of infinite
limits in integrals is a good approximation. For parabolic bands
$\upsilon_g = p_z/m$, and this relation is widely used in deriving
the well-known Fowler-Nordheim equation. In the case of carbon
nanotubes, two important differences from the generally accepted
consideration should be taken into account.

\begin{figure}
\centering
\includegraphics[width=4cm,  angle=270]{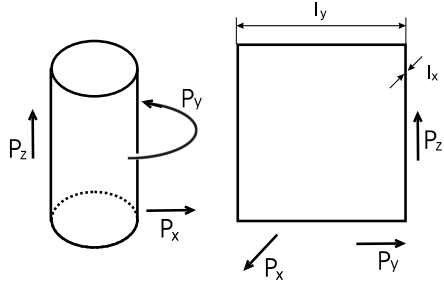}
\caption{Selected coordinate axes for a rolled (left) and unrolled (right)
nanotube. $l_y$ is the circumference of the nanotube,
$l_x$ is a thickness of the graphite layer.}
\label{figure1}
\end{figure}

First, an open SWNT has a finite small radius which results in
quantization of momentum. In this case, the corresponding
integrals in (\ref{main}) transform into sums. For example,
choosing the axes like shown in Fig.~\ref{figure1}
one has $\int f(p_i) d p_i  = \sum\limits_q f(q) h/l_i$ where
$i=x,y$, and $l_y$ is the circumference of the nanotube, $l_x$ the
thickness of the graphite layer. For a SWNT there exists only a
single layer in the $x$-direction and, accordingly, there is
exactly one term in the sum for $i=x$. The number of terms for
$i=y$ depends on the tube circumference $l_y$.

Second, the energy near the Fermi level for a single graphite
layer (graphene) is approximated by $\varepsilon = \pm \upsilon_F
\sqrt{p_y^2 + p_z^2}$, where $\upsilon_F$ is the Fermi
velocity~\cite{Wallace, Sloncz, McClure, Charlier}. The electrons
move in the $yz$ plane, so that the energy does not depend on
$p_x$. Since $p_y$ is quantized the energy $\varepsilon(p_y,p_z)$
turns out to be divided into a set of channels with
$\varepsilon(\hbar q, p_z)= \varepsilon^q(p_z)$ where $q$ takes
integer values. Therefore, the current density (\ref{main}) takes
the form
\begin{equation}\label{off}
j^{out} = \frac{2e}{h l_x l_y}\sum_q\int
f(\varepsilon^q)D(\varepsilon^q)d \varepsilon^q.
\end{equation}
As is known, the dispersion relation for carbon nanotubes depends
on their chirality (see, e.g.,~\cite{Dress}). For a chiral vector
$(m, n)$ it can be written as
\begin{equation}
\varepsilon^q =\pm \upsilon_F\left[{\left(h\frac{(m-n)/3 +
q}{l_y}\right)^2 + p_z^2}\right]^{1/2}. \label{eheh}
\end{equation}
Generally, there are two symmetric curves with a gap
$\varepsilon^q_g = 2\upsilon_F h((m-n)/3 + q)/l_y$. However, at
certain values of $m$, $n$ and $q$ the gap turns out to be zero
and one gets the linear dispersion relation. Therefore, at fixed
$q$ there exist one metallic branch and a set of semiconducting
branches for a SWNT with a given chirality.

As stated above, the condition $\upsilon_g > 0$ imposes
restrictions on the limits of integral in  (\ref{off}). In
addition, two approximations will be used. First, we consider the
zero-temperature limit when the Fermi-Dirac distribution becomes
the step function. Second, we suggest that the transmission
probability is given by the WKB approximation (see,
e.g.,~\cite{gp, Edgcombe2}) in the form
\begin{equation}\label{DD}
D(\varepsilon) = \exp\left\{-\frac{\zeta}{F}[
\phi^{3/2}\upsilon(y) - 3/2 \phi^{1/2}\varepsilon t(y)]\right\} =
b\exp(d\varepsilon)
\end{equation}
where $\zeta = 8 \pi(2 m_0)^{1/2}/3eh$, $y = (e
F/4\pi\varepsilon_0 )^{1/2}/\phi$, $F$ is the electric field,
$\phi$ the work function, $\varepsilon_0$ the dielectric
constant, and we denoted $b = \exp\left(-\zeta\phi^{3/2}
\upsilon(y)/F\right)$ and $d = 3\zeta\phi^{1/2} t(y)/2F$ for
convenience. The functions $\upsilon(y)$ and $t(y)$ describe a
deviation of the barrier from the triangle form due to image effects and
can be approximated by \cite{Haw}
\begin{equation}
\upsilon(y) \approx 1 - y^{1.69}, \qquad t(y) \approx  1 + 0.127
y^{1.69}.
\end{equation}
Now we are able to calculate the $q$th term in the sum
(\ref{off}). For the metallic branch the integration in
(\ref{off}) spreads from $-\infty$ to $0$. One obtains
\begin{equation}\label{term}
j_0 = \frac{4}{3}\frac{e}{h l_x l_y}\frac{F}{\zeta\phi^{1/2} t(y)
}\exp\left(-\frac{\zeta}{F}\phi^{3/2} \upsilon(y)\right)= \frac{2
e b}{h l_x l_y d}.
\end{equation}
For semiconducting branches the range of integration in
(\ref{off}) is ($-\infty$, $-\varepsilon_g^q/2$) and
\begin{equation}\label{eee}
j_q = j_0 \exp\left( -\frac{3}{2}\frac{\zeta\phi^{1/2} t(y)}{F}
\frac{\varepsilon_g^q}{2}\right) = j_0 \exp\left( - d
\frac{\varepsilon_g^q}{2}\right).
\end{equation}
Notice that the dispersion relation enters (\ref{eee}) only
through the gap. This agrees with the well-known fact that the
group velocity and the density of states are canceled in the
one-dimensional case \cite{gp}.

\begin{figure}[h]
\centering
\includegraphics[width=6cm,  angle=270]{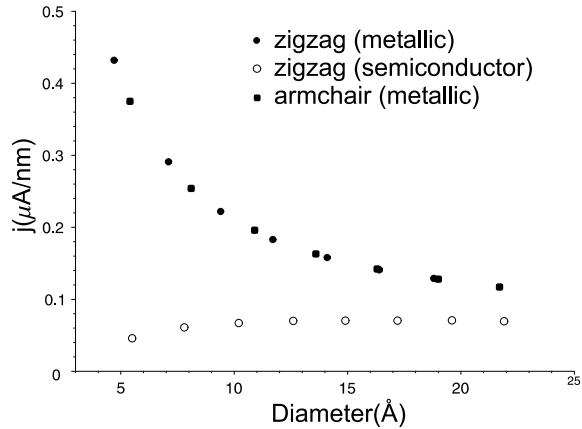}
\caption{Current densities vs the diameter of CNTs at the applied
field $F=8  \times 10^9$~V/m. The parameter set is $\phi = 4.7$
eV, $\upsilon_F = 0.83\times 10^6$~m/s, $a = 2.46$~\AA, $\zeta =
6.83 \times 10^9$ eV$^{-1/2}$ m$^{-1}$.} \label{fig:tube}
\end{figure}

The sum over all branches in (\ref{off}) gives the total FEC.
Fig.~\ref{fig:tube} shows the calculated emission current density,
which is a current divided by the circumference of a nanotube.
For $(m,n)$ SWNT the circumference is defined as $l_y = a\sqrt{m^2
+ mn + n^2}$ with $a$ being the lattice constant. In fact, the
main contribution to the sum in (\ref{off}) comes from the first
few terms corresponding to branches close to the Fermi level. This
is due to the exponential dependence of the FEC on the gap. For
metallic nanotubes, the leading term is $j_0$, so that
$j^{out}_{met}\sim 1/l_y$. In the case of semiconducting
nanotubes, the leading contribution comes from the $q$th term with
the smaller gap in (\ref{eee}) and, therefore, $j^{out}_{sem}\sim
1/[l_y\exp(1/l_y)] $. A similar behavior was found numerically
in~\cite{china1}. Moreover, comparing our results in
Fig.~\ref{fig:tube} with the exact numerical calculations
in~\cite{china1} one can find out a good qualitative agreement.
Notice that the quantitative difference is also not great and
varies from a few to ten percent depending on the diameter of the
nanotube. In comparison with~\cite{china1} our points in
Fig.~\ref{fig:tube} are situated slightly lower for metallic
nanotubes and slightly higher for semiconducting nanotubes. This
difference can be explained by at least two reasons. First, we
have used the simplified expression for the tunneling probability
in (\ref{DD}) where the image effects were approximated in a
standard way (see, e.g.,~\cite{gp, Edgcombe2}). Second, as
distinct from~\cite{china1} we consider the zero-temperature
limit.

\section{FEC of crystalline graphite}

\subsection{Noninteracting layers\label{nonint}}

\begin{figure}
\centering
\includegraphics[width=4.5cm,  angle=270]{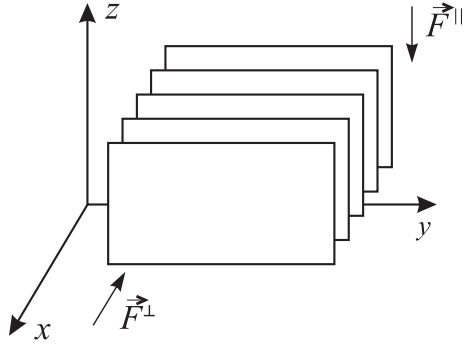}
\caption{The location of graphite layers with respect to the
electric field. The emission occurs in the direction opposite
to the electric field.} \label{fig:orient}
\end{figure}

In this section, we study a case of noninteracting graphite
layers. The layers are oriented as shown in Fig.~\ref{fig:orient}.
To calculate the FEC we will use the method of independent
channels described in the previous section. Namely, let us
consider the 2D graphite lattice with the Born-von Karman boundary
conditions applied in the $y$ direction. This gives the natural
quantization conditions. For layers of a large (infinite) size the
sum in (\ref{off}) can be replaced by the integral $\sum\limits_q
j_q = (l_y/h) \int j(p_y) d p_y$, so that finally one obtains
\begin{equation}\label{g-j}
j=\frac{8}{9}\frac{q}{h^2l_x}\frac{F^2}{\zeta^2\phi\upsilon_F
t^2(y) }\exp\left(-\frac{\zeta}{F}\phi^{3/2}\upsilon(y)\right),
\end{equation}
where the relation $\varepsilon_g(p_y) = 2\upsilon_F |p_y|$ is
taken into account. It is interesting to mention that this result
is very similar to the Fowler-Nordheim formula
\begin{equation}\label{f-n}
j^{FN} = \frac{16}{9}\frac{q m_0 \pi}{h^3}\frac{F^2}{\zeta^2\phi
t^2(y) }\exp\left(-\frac{\zeta}{F}\phi^{3/2}\upsilon(y)\right).
\end{equation}
Indeed, the exponents are exactly the same and the preexponential
factors differ only slightly. What is important, the
$F^2$-dependence is equal in both cases. For the interlayer
distance $l_x =  3.34$~\AA\ one can estimate
$j/j^{FN} = h/(2\upsilon_F\pi m_0 l_x) \sim 0.4$.

\subsection{$AAA$ stacking}

Generally, there are known three possible configurations of
crystalline graphite: $ABAB...$ stacking sequence of hexagonal
layers (Bernal structure), rhombohedral $ABCABC...$ stacking,
and $AAA...$ stacking when layers of carbon atoms are located directly
on top of each other~\cite{Charlier}. The $AAA$ stacking is called hypothetical because
it has not been observed yet in crystalline graphite. However, this
configuration is expected in disordered or pregraphitic
carbon~\cite{Charlier2}. In this paper, we consider the model of
$AAA$ stacking which is the simplest one and allows us to study
the effect of interlayer interaction on the emission properties
analytically.

Let us consider the interacting graphite layers oriented
parallelly to the electric field (see Fig.~\ref{fig:orient}). In
the framework of the $AAA$ model the interaction modifies the 3D
band structure near the Fermi level which can be written in a
simple analytical form \cite{Charlier}
\begin{equation}\label{3ds}
\varepsilon = 2 \alpha_1 \cos\left( \frac{c p_x}{\hbar}\right) \pm
\alpha_0 \frac{\sqrt{3} a }{2 \hbar}  \sqrt{p_y^2 + p_z^2}.
\end{equation}
Here the $\alpha_0$ parameter represents the interaction between
first-neighboring atoms in a layer, $\alpha_1$ is related to the
interaction between two atoms of the same projection on the $yz$
plane, from two neighboring layers, and $c$ is the interlayer
spacing. The influence of other parameters $\alpha_2$ and
$\alpha_3$ introduced in \cite{Charlier} is suggested to be
negligible and only linear terms in $\textbf{k}\cdot\textbf{p}$
perturbation expansion are taken into account. Actually, the
analysis in~\cite{Charlier} shows that the maximum effect of the
next-to-leading term (breaking the cylindrical symmetry) is of the
order of five percent. The upper sign in (\ref{3ds}) corresponds
to the conduction band, and the lower sign corresponds to the
valence band, $\alpha_0\sqrt{3}a/2 \hbar=\upsilon_F$.

\begin{figure}[h]
\centering
\includegraphics[width=6cm,  angle=270]{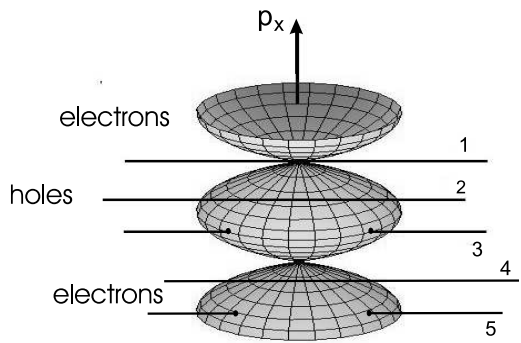}
\caption{Fermi surface of the simple hexagonal graphite. The central pocket
corresponds to the hole region, two half pockets correspond to the electron region.
The solid lines show five possible types of channels for emitting electrons.} \label{fig:fermi1}
\end{figure}

In compliance with (\ref{3ds}) Fig.~\ref{fig:fermi1} represents
the Fermi surface of the $AAA$ graphite. The Fermi surface is
composed of a hole pocket (the valence band, sigh "minus" in
(\ref{3ds})) and two half pockets of electrons (sign "plus" in
(\ref{3ds})). In our case, the emission occurs along the z-axis.
Generally, the possible values of the momentum of emitting
electrons with respect to the Fermi surface can be collected into
five different groups. We call them {\it independent channels}.
Solid lines in Fig.~\ref{fig:fermi1} indicate five possible types
of independent channels: (1) for an intermediate electron-hole
region, (2,3) for holes, and (4,5) for electrons. For finite-size
layers, quantization of momentum in the $xy$ plane occurs. In this
case, the spectrum is written as
\begin{equation}\label{3dsij}
\varepsilon^{ij} = 2 \alpha_1 \cos(c p_x^i/\hbar) \pm\upsilon_F
\sqrt{(p_y^{j})^2 + p_z^2},
\end{equation}
where $i,j$ are integer, and $p_x^i$ lies in the region $(-\pi
\hbar /c, \pi \hbar/c)$. The total FEC is a sum of all channels.
Let us consider these contributions separately.

\subsubsection{Hole region}

\begin{figure}[h]
\centering
\includegraphics[width=4.5 cm,  angle=270]{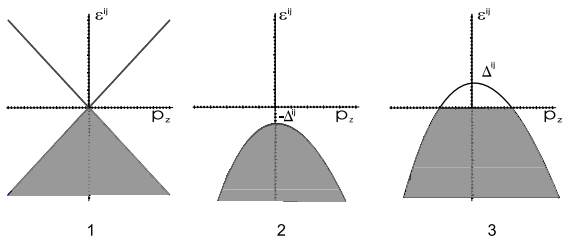} %{wood.bmp}
\caption{One-dimensional dispersion relations for the hole
region.} \label{holechannel}
\end{figure}

The hole region is defined as $-\pi\hbar/2c < p_x^i < \pi\hbar/2
c$. As is shown in Fig.~\ref{fig:fermi1} there are two types of
channels for the hole region and the channel 1 can be considered
as an intermediate case. Fig.~\ref{holechannel} shows all possible
one-dimensional  dispersion relations for this case.
The channel 3 crosses the Fermi surface at two points while the
channel 2 does not cross the Fermi surface. For each channel
$j^{ij} = (2e/h l_x l_y)\int f(\varepsilon^{i j})D(\varepsilon^{i
j})d\varepsilon^{ij}$. The integration here is over the occupied
states, and $\Delta^{i j}$ is the distance from the Fermi level to
the extremum point of the branch (see Fig.~\ref{holechannel}). For
the channel 1 $\Delta^{ij}=0$, and the current density is found to
be equal to $j_0$. Notice that there are only two channels of this
type. Analytically, the channel 2 is defined as $|p_y^{j}| >
\eta^i$, where $\eta^i=2\alpha_1 \cos(c p_x^i/\hbar)/\upsilon_F$.
For this channel one obtains
\begin{equation}\label{ij2j}
j^{i j} = j_0 \exp(-d \Delta^{ij}).
\end{equation}
Here $\Delta^{ij}$ takes the form $\Delta^{ij} =
\upsilon_F |p_y^{j}| - 2 \alpha_1\cos(cp_x^i/\hbar)$,
which can be easily found from (\ref{3dsij}). Replacing the sum
$\sum\limits_{ij} j^{ij}$ by the integral one gets
\begin{equation}
j^{\parallel}_2=
 \frac{2l_x
l_y}{h^2} \int\limits_{-h/4c}^{h/4c}d
p_x\int\limits_{\eta}^{\infty} j_0 \exp[-d\Delta(p_x, p_y)] d p_y
=\frac{2eb}{h^2d^2 c\upsilon_F}.
\end{equation}
For the channel 3 one has $|p_y^{j}|<\eta^i$ and $j^{i j}=j_0$.
Like before, one obtains
\begin{equation}
j^{\parallel}_3 = \frac{2l_x l_y}{h^2} \int\limits_{-h/4c}^{h/4c}d
p_x \int\limits_{0}^{\eta}j_0 d p_y = \frac{8\alpha_1 e b}{\pi h^2
dc\upsilon_F }.
\end{equation}

\subsubsection{Electron region}

\begin{figure}[h]
\centering
\includegraphics[width=4.5 cm,  angle=270]{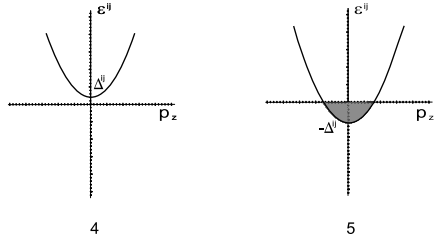}
\caption{One-dimensional dispersion relations for the electron
region} \label{electronch}
\end{figure}

The electron region is defined as $-\pi\hbar/ c < p_x^i <
-\pi\hbar/2 c$ and $\pi\hbar/2 c < p_x^i < \pi\hbar/ c$ or, taken
into account a periodicity of the Brillouin zone,  $\pi\hbar/2 c <
p_x^i < 3\pi\hbar/2 c$. There are two kinds of channels in the
electron region with spectra shown in Fig.~\ref{electronch}.
The channel 4 does not cross the Fermi surface and, therefore, the
current density turns out to be zero. For the channel 5 one has
$|p_y^{j}| < |\eta^i|$. This channel contains occupied states
below the Fermi level and, hence, there is nonzero contribution to
the FEC. One gets
\begin{equation}
j^{ij} = j_0 [1 - \exp(- d \Delta^{ij})],
\end{equation}
where $\Delta^{ij} = 2\alpha_1|\cos(c p_x^{i}/\hbar)| - \upsilon_F
|p_y^{j}|$ in accordance with (\ref{3dsij}). Finally,
\begin{eqnarray}
 j^{\parallel}_5 = \frac{2l_x l_y}{h^2} \int\limits _{h/4
c}^{3h/4c}d p_x\int\limits^{\eta}_{0} j_0\{1 - \exp[-d \Delta(p_x,
p_y)]\}d p_y \nonumber\\
 = \frac{2 eb[4 d \alpha_1/\pi + \bigl(\textbf{I}_0(2
d\alpha_1) - \textbf{L}_0(2 d \alpha_1) - 1\bigr) ]}{d^2 c
\upsilon_F h^2},&
\end{eqnarray}
where $\textbf{L}_0(x)$ is the modified Struve function, and
$\textbf{I}_0(x)$ is the modified Bessel function.

\subsubsection{Resulting FEC}

The total current density is found to be
\begin{equation}\label{result}
j^{\parallel}_{tot} = \sum\limits_{i=2}^5
j^{\parallel}_{i}=\frac{2 e b[8 d \alpha_1/\pi +  \textbf{I}_0(2
d\alpha_1) - \textbf{L}_0(2 d \alpha_1) ]}{h^2d^2 c \upsilon_F }.
\end{equation}
Notice that $j^{\parallel}_{tot}$ reduces to (\ref{g-j}) for
$\alpha_1 = 0$ .

\subsection{$AAA$ stacking: perpendicular electric field}

\begin{figure}[h]
\centering
\includegraphics[width=6cm,  angle=270]{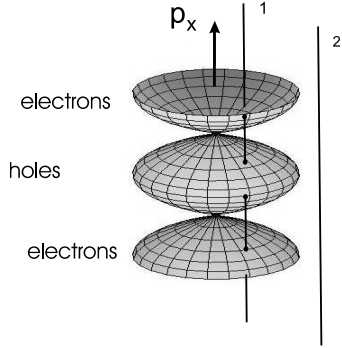}
\caption{Fermi surface of the simple hexagonal graphite. The solid
lines show two possible types of channels for electrons emitting in the
$x$-direction.} \label{channelsnormal}
\end{figure}

Let us consider interacting graphite layers placed normally to the
electric field (see Fig.~\ref{fig:orient}). This situation differs
markedly from the previous case. Let us denote $\rho=\sqrt{p_y^2+
p_z^2}$ in the dispersion relation in (\ref{3ds}). Quantization of
momentum results in the replacement $\rho\rightarrow\rho^{ij}$.
There are only two types of channels in this case (see
Fig.~\ref{channelsnormal}).
As before, let us consider them separately.

\subsubsection{Hole region}

\begin{figure}[h]
\centering
\includegraphics[width=4.5 cm,  angle=270]{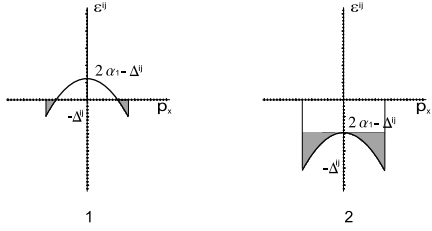}
\caption{One-dimensional dispersion relations for the hole
region.} \label{fig:holech}
\end{figure}

The hole region is defined by $-\pi\hbar/2c < p_x < \pi\hbar/2 c$.
There are two kinds of channels in the hole region with spectra
shown in Fig.~\ref{fig:holech}.
The channel 1 is defined by $\rho^{ij} < 2 \alpha_1/\upsilon_F$.
The current density reads
\begin{equation}
j^{ij} = j_0 [1 - \exp(-d \Delta^{ij})],
\end{equation}
where $\Delta^{ij}=\upsilon_F\rho^{ij}$. One obtains
\begin{eqnarray}
 j^{\perp}_1 = \frac{l_y l_z}{h^2} \int\limits_{0}^{2
\alpha_1/\upsilon_F} j_0 \{1 - \exp[-d \Delta(\rho)]\} 2 \pi \rho d \rho \\
\nonumber = \frac{4 \pi e b}{(d h)^3
\upsilon_F^2}\left(\frac{(2\alpha_1 d)^2}{2} - 1 + \exp(-
2\alpha_1 d)(2\alpha_1 d + 1) \right).
\end{eqnarray}
The channel 2 is defined by $\rho^{ij} > 2 \alpha_1/\upsilon_F$,
and
\begin{equation}
j^{ij} = j_0(\exp(2 \alpha_1 d) - 1) \exp(-d \Delta^{ij}).
\end{equation}
Finally,
\begin{eqnarray}
 j^{\perp}_2 = \frac{l_y l_z}{h^2} \int\limits_{0}^{2
\alpha_1/\upsilon_F} j_0 (\exp(2 \alpha_1 d) - 1) \exp[-d
\Delta(\rho)] 2\pi\rho d\rho
\\ \nonumber =\frac{4 \pi e b(2\alpha_1 d + 1)}{(d h)^3 \upsilon_F^2}
\exp(-2\alpha_1 d)(\exp(2\alpha_1 d) - 1).
\end{eqnarray}

\subsubsection{Electron region}

\begin{figure}[h]
\centering
\includegraphics[width=4.5cm,  angle=270]{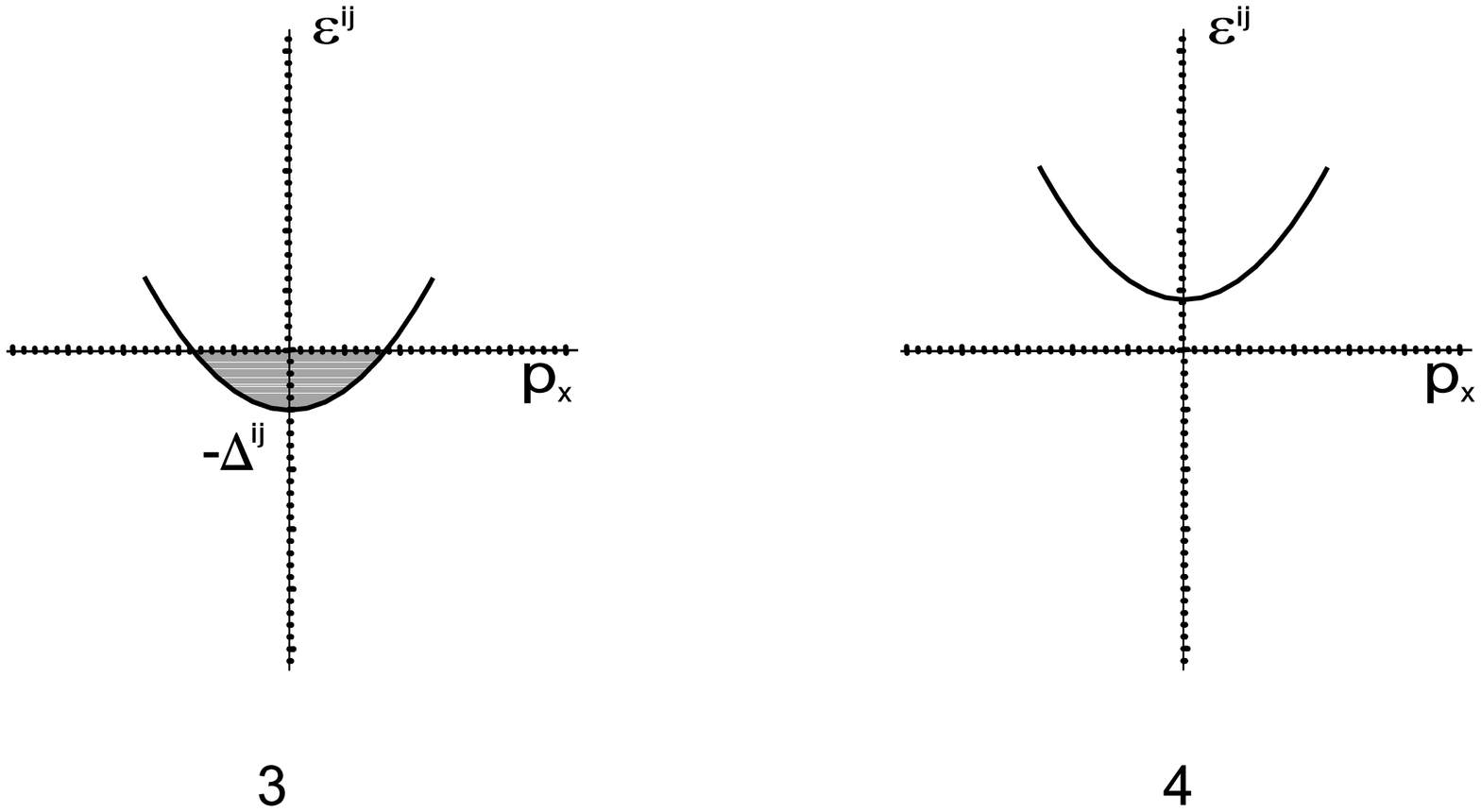}
\caption{One-dimensional dispersion relations for the electron
region.} \label{fig:electronch}
\end{figure}

In the electron region $\pi\hbar/2 c < p_x < 3\pi\hbar/2 c$. There
are also two kinds of channels in this region with spectra shown
in Fig.~\ref{fig:electronch}.
The channel 3 is equivalent to the channel 5 in the previous
section. It is defined by $\rho^{ij} < 2 \alpha_1/\upsilon_F$. The
current is
\begin{equation}
j^{ij} = j_0 [1 - \exp(- d \Delta^{ij})].
\end{equation}
Here $\Delta^{ij} = 2\alpha_1 - \upsilon_F \rho^{i j}$ (see
Fig.~\ref{fig:electronch} and (\ref{3dsij}). In this case,
\begin{eqnarray}
 j^{\perp}_4 = \frac{l_y l_z}{h^2} \int\limits_{0}^{2
\alpha_1/\upsilon_F}
j_0 \{1 - \exp[-d \Delta(\rho)]\} 2 \pi \rho d \rho\\
 =\frac{4 \pi e
b}{(d h)^3 \upsilon_F^2}\left(\frac{(2\alpha_1 d)^2}{2} -
2\alpha_1 d + \exp(2 \alpha_1 d) - 1 \right).
\end{eqnarray}
The channel 4 is defined by $\rho^{ij} > 2 \alpha_1/\upsilon_F$.
Since it does not cross the Fermi surface the current density is
equal to zero.

\subsubsection{Resulting FEC}

The total current density is the sum of all channels
\begin{equation}\label{result2}
j^{\perp}_{tot}=\frac{4 \pi e b}{(d h)^3
\upsilon_F^2}\left[(2\alpha_1 d)^2  + 1 - \exp(-2 \alpha_1
d)\right].
\end{equation}

\section{Discussion}

\begin{figure}[ht]
\centering
\includegraphics[width=6cm,  angle=270]{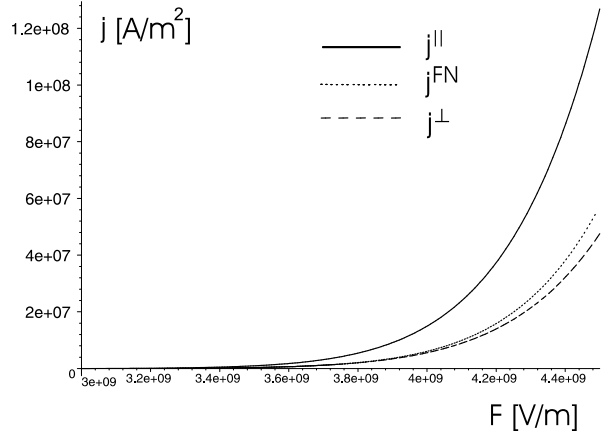}
\caption{The current densities vs electric field for graphite layers
placed parallelly ($j^\parallel$) and normally ($j^\perp$) to
the electric field with $\alpha_1=0.4$ eV, $\phi=5$ eV.
For comparison the Fawler-Nordheim curve ($j^{FN}$) is shown.} \label{fig:jf}
\end{figure}

\begin{figure}[h]
\centering
\includegraphics[width=6 cm,  angle=270]{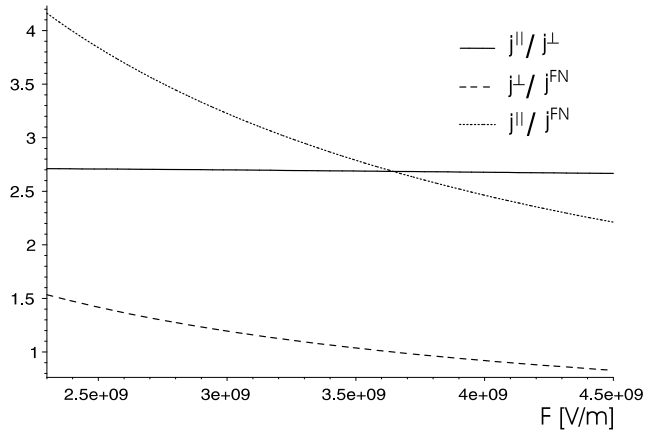}
\caption{Comparative curves for the current densities vs the electric field
characterizing the role of the preexponential factor.
$j^\parallel/j^{\perp}$ is almost a constant in the
considered interval of $F$.
}
\label{fig:all}
\end{figure}

Fig.~\ref{fig:jf} shows the calculated current densities as
functions of the applied electric field. For comparison, the
Fowler-Nordheim curve is drawn.
The most important difference comes from preexponential factors.
As is known, the Fowler-Nordheim theory gives $ j^{FN} \sim F^2b$
at all $F$. In accordance with (\ref{result}) the preexponential
factor has a different field-dependence. At small $F$ one obtains
$j^\parallel \sim Fb$. When $F$ increases (which means $2\alpha_1
d\rightarrow 0$) the current density comes to $j^\parallel \sim
F^2b$ and, finally, $j^\parallel/j^{FN}\rightarrow 0.4$ as was
shown in section \ref{nonint}. Indeed, at large $x$ the difference
$\textbf{I}_0(x)-\textbf{L}_0(x)$ tends to $2/(\pi x)$ while at
small $x$ it tends to $1$~\cite{Hand}. This is clearly seen in
Fig.~\ref{fig:all} where the comparative curves are demonstrated.
For $j^\perp$ we have a similar behavior. According to
(\ref{result2}), $j^\perp\sim \alpha_1^2 Fb$ at small $F$, and
$j^\perp\sim \alpha_1 F^2b$ at large $F$, so that $j^\perp/j^{FN}$
tends to a constant with increasing $F$. One can conclude that the
bigger is the electric field the lesser is the role of the
interlayer interaction. The anisotropy of the emission from the 3D
graphite is also shown in Fig.~\ref{fig:all}. As is seen,
$j^\parallel/j^{\perp}\sim 2.7$, that is almost a constant in the
considered interval of $F$. Therefore, we obtain three times
increase in FEC when graphite layers are oriented in parallel with
the electric field.

It is interesting to discuss the dependence of FEC from the
parameter $\alpha_1$ which characterizes the interlayer
interaction. It was found in \cite{Charlier2} that this parameter
is very sensitive to the interlayer distance.
Based on their results one can approximate
\begin{equation}
\alpha_1 = 18 x^2 - 0.4,
\end{equation}
where $x = (c - c^*)/c$ and $\alpha_1$ is measured in eV.
\begin{figure}
\centering
\includegraphics[width=6cm,  angle=270]{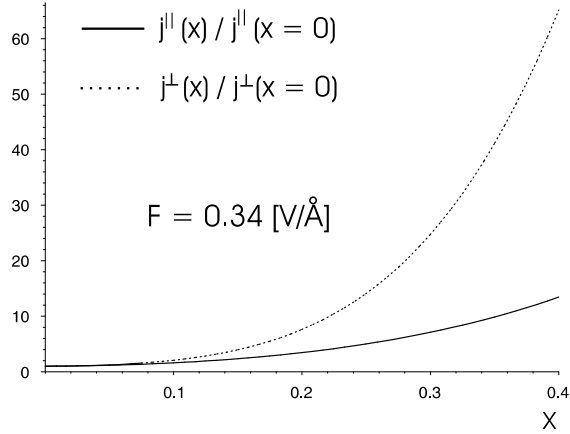}
\caption{Reduced current density vs the relative interlayer
distance $x = (c - c^*)/c$, $c = 3.34 \AA$ for two field orientations.
$j^\perp$ is markedly more sensitive to $x$ than $j^\parallel$.}
\label{interl}
\end{figure}
As is seen from Fig.~\ref{interl}, there is a strong dependence of
the FEC from the interlayer distance. The less is this distance
the more is the emission current. This is valid for both
orientations. It would be interesting to check this finding in
experiments with graphite crystals under pressure. Notice that
this result follows from the fact that the DOS at the Fermi level
(which is of the most importance in the emission process) is
determined by $\alpha_1$ (see \cite{Charlier}). As is seen from
Fig., \ref{interl}, $j^\perp$ is more sensitive to $x$ than
$j^\parallel$. Moreover, for $\alpha_1\rightarrow 0$ one has
$j^\perp\rightarrow 0$, which follows from the fact that the
movement of electrons between layers is suppressed in the absence
of the interlayer interaction.

\section*{Conclusion}

In conclusion, we have found that the band structure of the 3D
graphite has a marked impact on the field emission current.
Experimentally, the field emission from carbon materials was
studied in \cite{Obraztsov}. Unfortunately, the polycrystalline
carbon films used in experiment can not be properly described in the
framework of our approach because for this purpose we have
to consider a mixture between different crystalline structures.
In fact, the Fermi surface of the $ABAB$ structure of graphite
is found to be much more complex and, in particular, it does not possess
the cylindrical symmetry (see, e.g.,~\cite{McClure,Charlier2}).
In this case, our approach should be markedly modified.
Besides, many additional factors like the presence of a diamond-like phase
on the surface of samples and the absence of any information about
the local electric fields do not allow us to clarify the role
of the band structure in this case. Therefore, specific emission
experiments with graphite single crystals at different orientations
of the electric field would be of evident interest.

This work has been supported by the Russian Foundation for Basic
Research under grant No. 05-02-17721.

\end{document}